\documentstyle[preprint,aps,prl,epsf]{revtex}
\input{epsf}
\input epsf.sty
\begin{document}
\draft
\title{
Thermodynamic Phase Diagram of the Quantum Hall Skyrmion System 
}
\author{Kyungsun Moon}
\address{
Department of Physics, Yonsei University, Seoul 120-749, Korea\\
}
\author{Kieran Mullen}
\address{
Department of Physics and Astronomy, University of Oklahoma, Norman,
OK~~73019\\
}
\date{\today}
\maketitle

{\tightenlines
\begin{abstract}
We numerically study the
interacting quantum Hall skyrmion system based on the Chern-Simons
action.
By noticing that the action is invariant under global spin
rotations in the spin space with respect to the magnetic field direction,
we obtain the low-energy effective action for a many skyrmion system.
Performing extensive molecular dynamics simulations,
we establish the thermodynamic
phase diagram for a many skyrmion system.
\end{abstract}
}


\narrowtext

The application of a strong magnetic field 
to a two dimensional electron gas
``quenches'' the kinetic energy of the electrons - removing it from
consideration by constraining them to cyclotron motion.
This allows other 
physical effects such
as interactions and correlations to dominate.\cite{{prange}}
When the number of electrons, $N_e$, is an integer multiple of the number
of flux quanta, $N_\phi$, the groundstate of the system is a set of
completely filled Landau levels. 
If we define the filling factor $\nu\equiv N_e/N_\phi$, then for $\nu=1$,
$1/3$ and $1/5$ the groundstate is fully spin polarized.  This polarization
arises not simply from the Zeeman splitting but rather from interactions;
indeed it persists even in the limit of zero Zeeman
coupling.\cite{{leekane},{sondhia},{kmoona}}
The system  minimizes its Coulomb energy by minimizing
the overlap of  electrons in the many-body wavefunction.  
When the electrons all have the same spin, 
the amplitude for two electrons to be at the same spot must vanish
due to the exclusion principle, thereby reducing this electrostatic energy.

Near $\nu=1$ the
low energy excitations of the system are distortions of the spin
orientation, which are coupled to the local charge density. The system
arranges spins of adjacent electrons so that they are {\em nearly}
parallel, forming smooth distortions or textures in the spin field.
This can be
described by a non-linear sigma
model,\cite{{sondhia},{kmoona}}
   which predicts localized charged   
excitations\cite{{leekane},{sondhia},{kmoona},{zeeC},{abolfathA}}
consisting of dimples in the spin field
called skyrmions.\cite{{skyA},{skyB},{smgrev}}
  It follows
 that if one moves away from $\nu=1$ the groundstate
will include such textures as (possibly pinned) quasiparticles.  This has
been verified experimentally by NMR measurements of the spin
polarization.\cite{{NMRa},{NMRb}}    Other experiments such as optical
absorption\cite{{optical}}
and transport\cite{{transport}} are consistent with this picture.

At higher
densities (i.e. further away from $\nu=1$)  these quasiparticles
themselves start to interact with each other.  Each skyrmion has an
overall U(1) symmetry corresponding to rotation about the
z-axis.\cite{{skycrysHD}}
When two are brought near each other, the lowest energy configuration is to
align themselves anti-ferromagnetically.  When many skyrmions interact,
there is a competition between the Coulomb interaction, which favors a
triangular lattice, and this antiferromagnetic XY interaction, which favors
a bipartite one such as a square lattice. The resulting phase
diagram is quite rich, and has been investigated
theoretically.\cite{{skycrys},{crysB},{crysC}}

While the skyrmion lattice has not been seen
directly, some experiments are suggestive. NMR measurements
display a dramatic increase in the nuclear spin relaxation rate as a
function of $|\nu-1|$,\cite{{NMRb}} and
the low temperature heat capacity of multilayer
2DEG samples has a dramatic peak
that may be related to the melting of a skyrmion lattice.\cite{{cvA}}
This anomalous, large ($\times 10^5$)
peak in the  specific heat in GaAs quantum wells is postulated to come
from a coupling of {\em all} of the nuclear magnetic moments of all the
atoms
adjacent to the quantum well, due to the symmetry
breaking field of the Skyrme crystal.

In this paper we numerically
study the phase diagram of the Skyrme crystal.
Elsewhere we have derived a low-energy effective action,
for the many body skyrmion system.\cite{{effact}}  This effective
action accurately describes the renormalization due to the higher Landau
level mixing and is shown to be accurate for large and small
skyrmion sizes.  
Here we use this action to variationally
determine a two-body interaction potential between skyrmions.
We then use this two-body interaction in a 
molecular dynamics (MD) simulation method
to study the thermodynamic phase diagram
of a many body interacting skyrmion system. 
Unlike earlier studies, we do {\em not} presume an {\em a priori} lattice
structure.  We incorporate the full, long range nature of the interaction
allowing for competition between incompatible lattice structures and the
possibility of frustration.  
At zero temperature, we find that the triangular lattice is formed below 
$\nu_1\cong 1.025$ and
the square lattice becomes stable above $\nu_1$. At high filling 
factor, the square Skyrme crystal will melt due to strong quantum 
fluctuations.\cite{skycrysHD} 
Finite temperature simulation shows that at $\nu\cong 1.1$,  
the square Skyrme crystal melts at about $T_m\cong
0.035 \rho_s$. For the appropriate choice of experimental parameter, {\em i.e.}
$\rho_s\cong 1.5 {\rm K}$,  
$T_m$ is estimated to be $50{\rm mK}$, which  
is very close to the value given by experiment.\cite{cvA}
%We do not find a glassy or frustrated phase.

We begin with the effective action ${\cal S}_E$ for many skyrmion 
system\cite{{effact}} 
\begin{eqnarray}
{\cal S}_E [{\bf m}]&=& {1\over 2}\sum_{{\bf k},\omega} V_{\rm eff}(k)
|J_0^s|^2 + \sum_{{\bf k},\omega}
{1\over 2\kappa} |{\bf J}^s|^2  \nonumber\\ 
&+& \frac {\rho_s}{2} \int d{\bf r} \,\, (\nabla {\bf m})^2 
+ {1\over 2}g\mu B\,
m_z \nonumber\\
&+&{i\over \nu}\sum_{{\bf k},\omega} {\bf A^{(0)}}\cdot {\bf J}^s 
-{2\pi\over \nu}\sum_{{\bf k},\omega} \frac {1}
{k^2} J_0^s {\hat z}\cdot
{\bf k}\times {\bf J}^s  
\label{baction}
\end{eqnarray}
where $\kappa={\bar \rho}/m^*$, $V_{\rm eff}(k)=V(k)/(1+bk)$
is a screened Coulomb interaction between
skyrmions, $V(k)=2\pi e^2/\epsilon k$ is a bare interaction,  $\epsilon$ is
the dielectric constant,
$b=\ell^2/a_B$ with $a_B=(\hbar^2\epsilon/m^* e^2)$, 
$\ell=(\hbar c/|e|B)^{1/2}$ is the magnetic length, and  
$\rho_s$ is the spin stiffness $e^2/16\sqrt{2\pi}\epsilon\ell$.
The skyrmion {\em three}-vector $J_\mu^s$ is related to ${\bf m}({\bf r})$,
the local orientation of the spin texture via:
\begin{equation}
J^s_{\mu} ={\nu\over 8\pi}
\epsilon_{\mu\nu\lambda} (\partial_\nu {\bf m}\times
\partial_\lambda {\bf m})\cdot {\bf m} .  \label{eqjmu}
\end{equation}
By noticing that the presence of a finite Zeeman gap makes
the skyrmion action invariant
under global spin rotations by an angle $\phi$ in the
${\rm XY}$-spin space\cite{{skycrysHD}},
we chose to describe a skyrmion as a point particle with
charge $q_{s}=\nu e$ and
phase $\phi$.
The Lagrangian for a many skyrmion system can then be simplified
\begin{eqnarray}
{\cal L}[{\bf R}_i,\phi_i]&=& {1\over 2} \sum_{i}\left\{ M_{s}\left
(\frac {d{\bf R}_i} {dt}\right)^2 + I_{s}
\left (\frac {d\phi_i}{dt}\right)^2 \right\} \nonumber\\
&-&{1\over 2}\sum_{i\ne j} V({\bf R}_i-{\bf R}_j,\phi_i-\phi_j) \nonumber\\ 
&+& {i\over \nu}\sum_{\bf k} {\bf A^{(0)}}\cdot {\bf J}^s 
-{2\pi\over \nu}\sum_{\bf k} \frac {1}
{k^2} J_0^s {\hat z}\cdot
{\bf k}\times {\bf J}^s .
\label{Lagra}
\end{eqnarray}
The skyrmion transport mass $M_{s}$ and the moment of inertia
$I_{s}$ are given by the following relations
\begin{equation}
M_{s}={2\pi m^*\over \nu} \int_{\cal S} d{\bf r}\ell^2 \left|J^s_0\right|^2
\end{equation}
\begin{equation}
I_{s}={2\pi m^*\ell^2\over \nu}\sum_{i=x,y}\int_{\cal S} d{\bf r} \left|
{\nu\over 4\pi} ({\bf m}\times \sigma_y {\bf m}_\perp)\cdot
\partial_i {\bf m}\right|^2
\end{equation}
where
${\cal S}$ represents an area occupied by a single skyrmion
and $\sigma_y$ is a Pauli spin matrix. The function $V({\bf R},\chi)$ is
the
interaction between two skyrmions, where $R=|{\bf R}_i-{\bf R}_j|$ is the 
relative distance and $\chi=(\phi_i-\phi_j)$ relative spin orientations. 
The last two terms in Eq.(\ref{Lagra})
take into account the exchange statistics of a skyrmion.\cite{effact}
For small $\chi$, the two-body interaction can be decomposed
into the following suggestive form
\begin{equation}
V[{\bf R},\chi]=V_0(R)+V_A (R)[1+\cos(\chi)] .
\label{eq:potential}
\end{equation}
At large distances, $V_0 (R)$ goes like $q_{s}^2/\epsilon R$ as
a bare Coulomb repulsion between two point charges and
$V_A (R)$ decays exponentially due to a finite Zeeman gap.
Note that the angle dependent interaction $V_A (R)$ favors
an antiferromagnetic spin alignment, {\em i.e.} $\chi=\pi$,
trying to form a
{\em bipartite} lattice,
while $V_0 (R)$ prefers a triangular lattice
which does {\em not} belong to {\em bipartite} lattices. 
%\begin{figure}
%\epsfxsize=2.5in
%\epsfysize=2.5in
%\epsffile{arrowfield.eps}
%\vskip 0.5cm
%\epsfxsize=2.5in
%\epsfysize=2.5in
%\epsffile{charge.eps}
%\vskip 0.5cm
%\caption{A typical spin texture and its associated
%charge distribution for two skyrmions with a
%relative twist in their spin field.  
%The charges are set 1.5$\ell$ apart,
%$\xi=0.22$, and $\theta_0=\pi/2$.  Note the slight charge ridge between the
%two main charge centers; this ridge shifts as a function of $\theta_0$.
%}
%\label{texture}
%\end{figure}
This two-body interaction is difficult to calculate variationally: 
A configuration of two skyrmions 
must be allowed to relax {\it without} allowing the skyrmions to
migrate apart.  We chose an ansatz  that 
fixed the relative angle between two skyrmions to be $\chi$ and the distance
to be $R$, and then calculated 
the minimum energy configurations numerically.  Our variational form is:
\begin{equation}
\omega (z,z^*)=\frac {z^2-\left({R\over 2}\right)^2}{\lambda^2} \left[
\frac {2 \,\, e^{i\theta (x)/2}}{e^{-|z-R/2|/2\xi}+e^{-|z+R/2|/2\xi }}\right]
\end{equation}
where $z=x+iy$ denotes the position vector, $\xi=(2\pi/E_z)^{1/2}$
with $E_z=(1/2)g\mu B/\rho_s$,
and $\lambda$ is a variational
parameter
which controls the extent to which the spin fields are deformed.
The spin fields $\bf {m({\bf r})}$ are related to the complex field
$\omega(z,z^*)$ via the following relation: $m_x+i m_y=2\omega/(1+|\omega|^2)$.
\cite{skyA}
%KMOON(3.22)  
%In Fig.(\ref{texture}) we show a sample texture for
%two skyrmions with a relative twist of $\pi/2$, and the corresponding
%charge distribution associated with the texture.  
%%KMOON
When the distance $R$ is much larger than $\xi$, 
our variational ansatz describes two nearly free n=1 skyrmions, and
near $z=R/2$, $\omega (z,z^*)\cong [(z-R/2)/\lambda_s] \, e^{|z-R/2|/2\xi}$ 
with $\lambda_s=\lambda^2/(2R)$ and $\theta(x)$ chosen to be $0$. 
The variational energy of an n=1 skyrmion based on the above   
ansatz yields $1.789$ in the units of $4\pi\rho_s$ and    
$\lambda_s=2.34\ell$, which is remarkably close to  
the exact value $1.788$ from the field theoretic action.\cite{effact}
In the opposite limit of $R=0$, it depicts an n=2 skyrmion, whose variational 
energy is calculated to be $3.939$, which is
very close to the exact value of $3.913$ in the units of $4\pi\rho_s$,
which lends a strong credence to our variational ansatz.
%%%
By minimizing the action in Eq.(\ref{baction}) for the proposed two skyrmion
ansatz with $\theta (x)=0$, the two-body interaction $V_0 (R)$ is calculated. 
%and shown in Fig.(\ref{datfit}). 
In order to calculate the angle-dependent interaction $V_A (R)$,
we adjust $\lambda$ to 
minimize the energy for the following ansatz for the 
gradual spin twist $\theta (x)$, which   
rotates $\bf {m}$ with respect to the $\hat {z}$ axis:   
$$\theta (x)= \theta_0 \tanh(2x/R)$$
which uniformly inserts twist between the two skyrmions.  
%A sample spin 
%texture for two skyrmions, and the corresponding charge distribution is
%shown in Fig.(\ref{texture}).
Here $\pi-\theta_0$ is the relative phase difference $\chi$ between 
two skyrmions.
This ansatz does not change the Zeeman energy, 
since $m_z$ stays the same.
In Fig.(\ref{surfpot}) we plot the interaction 
strengths $V (R,\pi-\theta_0)$    
as a function of distance $R$ and relative angle $\theta_0$ between 
the textures.
We can break this into two parts as described
in Eq.(\ref{eq:potential}) and 
the interaction profile thus obtained is given by 
\begin{equation}
V_0 (R)\cong 3.578+\frac {0.35 + 0.175 R}{1 + 0.0213 R + 0.0147 R^2} 
\end{equation}
\begin{equation}
V_A (R)\cong 0.629 \, e^{-0.434 R}
\label{intA}
\end{equation}
where $V_0$ and $V_A$ are in the units of $4\pi\rho_s$, 
$R$ in the unit of $\ell$.
%%KMOON (Aug. 18) 
One can notice that $V_0 (R)$ bends over to a certain finite  
value as $R$ decreases due to the soft core nature of skyrmion and 
$\partial^2 V_0/\partial R^2$ becomes negative 
at $\nu_c\cong 1.2$. This leads to a complex pattern at filling factors around 
and above $\nu_c$\cite{Long}, which will most likely be preempted by
quantum melting.\cite{skycrysHD}    
%%%
%At low skyrmion densities, one can assume that the skyrmion can be considered
%to be a {\em soft} core boson neglecting the {\em rare} exchange event.
%\begin{figure}
%\epsfxsize=2.5in
%\epsfysize=2.5in
%\epsffile{datfit.eps}
%\vskip 0.5cm
%\caption{Numerically calculated interaction potential for two skyrmions;
%the points are data from variational calculations, and the curve is a
%numerical fit to this data.  The interaction potential matches well with
%exact  results in the limits $R\to 0$ and $R\to \infty$.
%}
%\label{datfit}
%\end{figure}
Based on the interactions obtained above, we employ overdamped molecular  
dynamics simulations at finite temperature  
\begin{eqnarray}
&\eta_R& \frac {d{\bf R}_i}{dt}=-\sum_{j\ne i} \partial_{{\bf R}_i}
V_0({\bf R}_{ij}) \nonumber\\
& & \,\,\; \; \; \; \; \; \; \; \; -\sum_{j\ne i} 
\partial_{{\bf R}_i} V_A({\bf R}_{ij})
[1-\cos(\phi_{ij}-\pi)] + {\vec \xi}^R_i (t) \nonumber\\ 
&\eta_\phi& \frac {d\phi_i}{dt}=-\sum_{j\ne i} V_A({\bf R}_{ij})
\sin(\phi_{ij}-\pi) + \xi^{\phi}_i (t)
\end{eqnarray}
\begin{equation}
\langle \xi^{\alpha}_i (t)\xi^{\alpha}_j (t^\prime)\rangle
= 2\eta_{\alpha} T\delta_{ij}\delta(t-t^\prime) .  
\end{equation}
Here $\eta_\alpha$ is the damping parameter with $\alpha=R,\phi$,
${\bf R}_{ij}={\bf R}_{i}-{\bf R}_{j}$, ${\phi}_{ij}={\phi}_{i}-{\phi}_{j}$,
and $\xi^{\alpha}_i (t)$ is a Gaussian white noise which represents  
the thermal fluctuations following the fluctuation-dissipation
theorem. It can also help the system to reach
the {\em true} ground state by providing a random noise.   

The number of skyrmions $N$ was fixed to be $576$ and the mean particle 
spacing $R_s$ varies with $\nu$ as $(2\pi/(\nu-1))^{1/2}$    
assuming the square lattice structure.   
Now we construct the phase diagram in the filling factor-temperature plane.  
The initial configurations are chosen randomly. We let the MD simulations
evolve with time, make sure that the system reaches the equilibrium state,
then freeze the skyrmion configuration and measure the static structure factor
$S({\bf k})$ 
\begin{equation}
S({\bf k})=  \; {1\over N^2}
\sum_{i,j} e^{i {\bf k}\cdot ({\bf R}_i (t)-{\bf R}_j (t))}.
\end{equation}
By measuring $S({\bf k})$ and visually checking 
the snap shots, we identify the corresponding phases. 
First we have studied the zero temperature phase diagram. 
Insets of Fig.(\ref{phasediagram}) demonstrate $S({\bf k})$ at $\nu=1.02$ 
and $1.07$ 
exhibiting the six and four-fold peaks, as expected for the triangular and
square lattice structures, respectively. 
At zero temperature, with the increase of filling factor away from $\nu=1$,    
the system initially forms 
a triangular lattice and above a critical value of $\nu_1\cong 1.025$
a square lattice. At higher densities, the square Skyrme crystal will 
melt due to strong quantum fluctuations.\cite{skycrysHD,Long} 
Subsequently, we investigate the melting lines for the various Skyrme crystals
via finite temperature MD simulations.  
The melting phase boundary can be established by measuring the peak heights of
static 
structure factor $S({\bf K}_0)$ as a function of temperature and  
${\bf K}_0=(2\pi,2\pi)$ for the square Skyrme crystal. 
The Lindemann criterion says that the solid melts when the thermal 
fluctuations of the equilibrium positions are about $15 \%$ of the mean 
particle spacing $R_s$, that is, $\sqrt{<{\bf u}^2>}\cong c_L R_s$ with 
$c_L\cong 0.15$.   
Assuming Gaussian fluctuations, we obtain that 
$S({\bf K}_0)\cong e^{-{\bf K}_0^2 <{\bf u}^2>/2}\cong 1/3$.\cite{Vinokur}
Thus when the peak height of $S({\bf K}_0)$ reaches about $1/3$, we consider
the crystal having melted. 
%\begin{figure}
%\epsfxsize=2.5in
%\epsfysize=2.5in
%\epsffile{PUB_Laue1.02.eps}
%\epsfxsize=2.5in
%\epsfysize=2.5in
%\epsffile{PUB_Laue1.07.eps}
%\vskip 0.5cm
%\caption{Static structure factor $S({\bf k})$: 
%(a) At filling factor $\nu=1.02$,  
%six-fold peaks appear at the corresponding points to the triangular
%lattice. (b) At filling factor $\nu=1.07$,  
%four-fold peaks appear at the corresponding points to the square lattice.
%}
%\label{triSF}
%\end{figure}
In Fig.(\ref{phasediagram}), 
we have plotted the melting phase boundary and noticed that 
at around $\nu=1.1$, the melting temperature is about $0.035 \rho_s$.
The experimental value of $\rho_s$ is about $1.5{\rm K}$ yielding $T_m\cong
50 {\rm mK}$, which is very close to the experimentally suggestive melting 
temperature.  
%
%Kieran
%We need to check the experimental value of \rho_s from the paper
%  Yes...  Not done this yet.  Next draft.
%
The spin structure factor $S_{\phi}({\bf k})$ has also been measured 
\begin{equation}
S_{\phi}({\bf k})=  \; {1\over N^2}
\sum_{i,j} e^{i {\bf k}\cdot ({\bf R}_i (t)-{\bf R}_j (t))}  
e^{i (\phi_i (t)-\phi_j (t))}.
\end{equation}
Within the square Skyrme crystal phase, 
our simulations clearly demonstrate that 
the system possesses an antiferromagnetic spin order by exhibiting the
peaks of $S_{\phi}({\bf k})$ at ${\bf k}=(\pi,\pi)$.\cite{Long}    
We note that although it is possible that there could be a regime where the
triangular and square lattices could compete to produce a  
disordered phase, we do not observe this with our model using the
experimentally relevant parameters. In order to clarify this interesting
possibility, further studies including careful finite size scaling analysis  
is required.\cite{Long}   
In this work, we do not
treat the terms in the action that guarantee the fermionic nature of
skyrmions,  
nor do we include any quantum dynamics in the MD simulation.\cite{Fertig}   
Finally, we ignore any three-body interactions, treating the
spin texture  action as a set of pairwise interactions between skyrmions.

In summary, we have determined an approximate two body interaction potential
for skyrmions 
from an effective action via a variational calculation.  Using this
potential we have performed molecular dynamics simulations that have
determined the phase diagram of the skryme system without having to posit a
crystal structure.  We find that the melting temperature within our model
is close to the specific heat peaks seen in proposed skyrme crystal
experiments, which gives a strong support to the very existence 
of skyrme crystal.  
%In the future we plan to include fermionic effects, and to 
%examine the stability of crystal structures for other systems.

\acknowledgements

It is our great pleasure to acknowledge S.M. Girvin, H. Fertig, and D.H. Lee
for useful discussions. 
K. Moon wishes to acknowledge the financial support of the Korea Research
Foundation made in the program year of 1998.
This work was supported in part by Yonsei University Research Fund of 1998 
and NSF grant DMR-9502555.

\begin{figure}
%\epsfxsize=3in
%\epsfysize=3in
%\epsffile{SkyrFig1.eps}
%\vskip 0.5cm
\caption{The two-body 
interaction potential as a function of $R$ and $\theta_0$, as 
determined by variational calculations of
the total energy.  The curve with $\theta_0=0$ represents $V_0 (R)$.
Note the steep dependence of the energy as a
function of twist.
}
\label{surfpot}
\end{figure}

\begin{figure}
%\epsfxsize=3.5in
%\epsfysize=3.5in
%\epsffile{SkyrFig2.eps}
%\vskip 0.5cm
\caption{Thermodynamic phase diagram of a manybody skyrmion system:
Below $\nu_1\sim 1.025$, triangular lattice is stable. Above $\nu_1$,
square lattice is formed, which melts at around $0.035 \rho_s$ for $\nu=1.1$.
At some higher density, square skyrme crystal will melt due to 
quantum fluctuations.  
Insets show static structure factor $S({\bf k})$: 
At filling factor $\nu=1.02$,  
six-fold peaks appear at the corresponding points to the triangular
lattice. At filling factor $\nu=1.07$,  
four-fold peaks appear at the corresponding points to the square lattice.
}
\label{phasediagram}
\end{figure}

%\begin{figure}
%\epsfxsize=3.in
%\epsfysize=3.in
%\epsffile{TwoSkyrm.eps} 
%\vskip 1cm
%\caption{Interaction profile of two skyrmions as a function of relative 
%distance $r/R_s$ with $R_s=(2\pi/(\nu-1))^{1/2}\ell$. 
%Here the solid line stands for $V_0 (R)$ and the dashed line
%$V_A (R)$. 
%}
%\label{twoskyrm}
%\end{figure}

%\begin{figure}
%\epsfxsize=3.in
%\epsfysize=3.in
%\epsffile{PUB_fill1.02.eps}
%\vskip 1cm
%\caption{Snap shot of triangular skyrme crystal at filling factor $\nu=1.02$.
%}
%\label{fig2_1}
%\end{figure}

%\begin{figure}
%\epsfxsize=3.in
%\epsfysize=3.in
%\epsffile{PUB_fill1.07.eps}
%\vskip 1cm
%\caption{Snap shot of square Skyrme crystal at filling factor $\nu=1.07$.
%}
%\label{fig3_1}
%\end{figure}

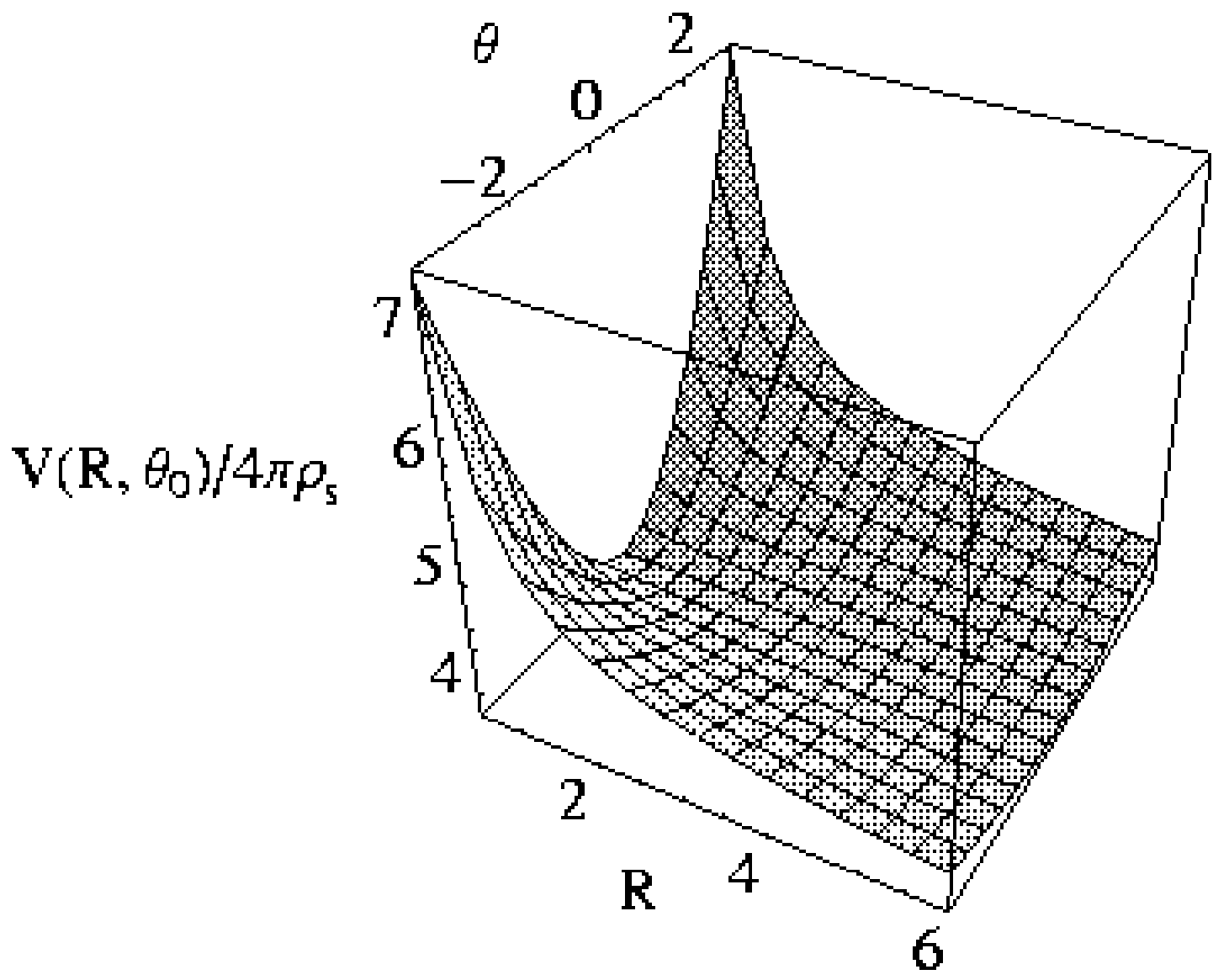
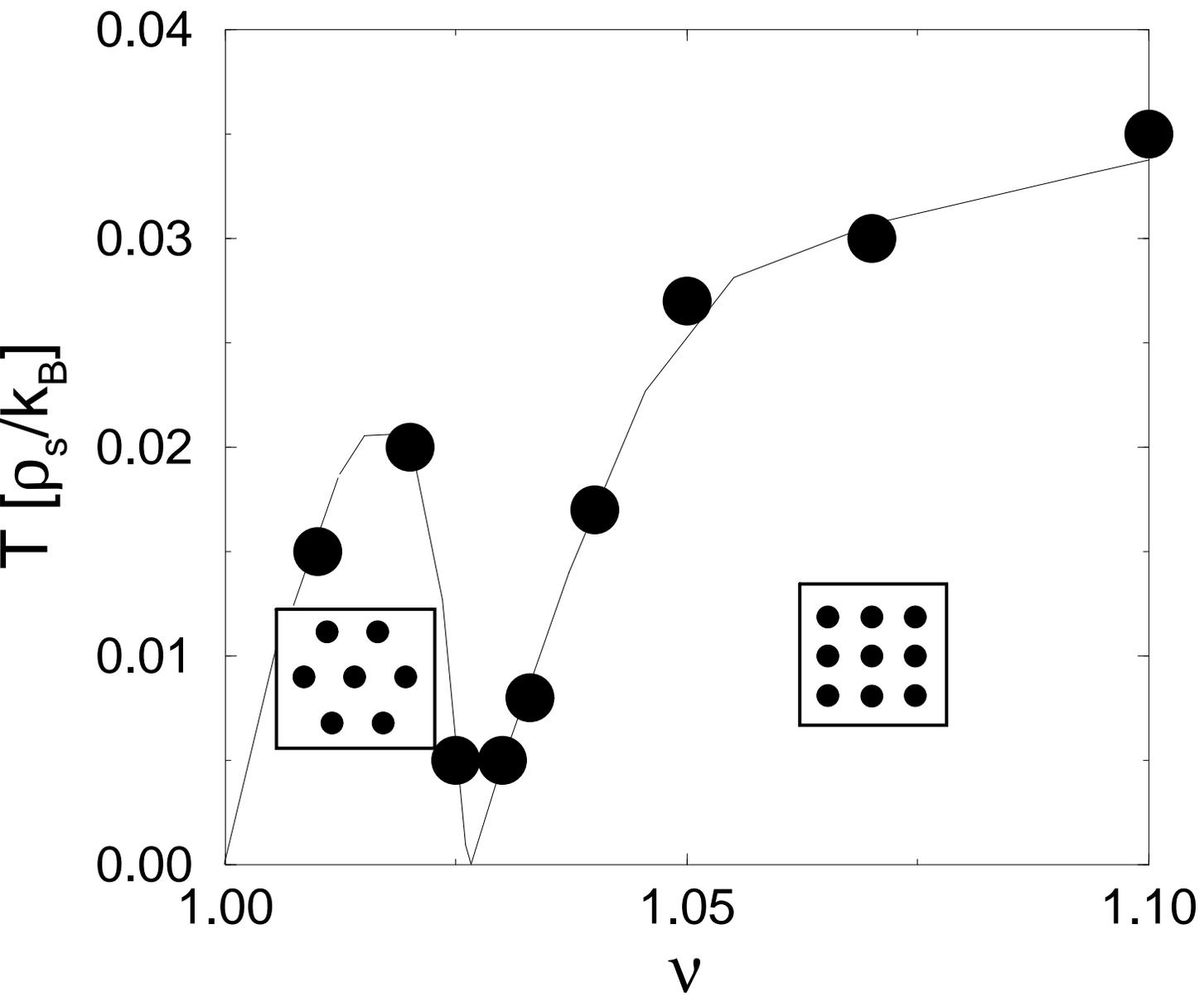
\end{document}